# Daisy: Data analysis integrated software system for X-ray experiments

*Yu Hu*[1], *Ling Li*[1,2], *Haolai Tian*[1,2,3*], *Zhibing Liu*[1,2], *Qiulan Huang*[1,2], *Yi Zhang*[1], *Hao Hu*[1], *Fazhi Qi*[1,2,#]

[1]Computing Center, Institute of High Energy Physics, Beijing, 100049, China.
[2]University of Chinese Academy of Sciences, Beijing 100049, China
[3]Spallation Neutron Source Science Center, Dongguan 523803, China

**Abstract.** Daisy (Data Analysis Integrated Software System) has been designed for the analysis and visualization of the X-ray experiments. To address an extensive range of Chinese radiation facilities community's requirements from purely algorithmic problems to scientific computing infrastructure, Daisy sets up a cloud-native platform to support on-site data analysis services with fast feedback and interaction. The plugs-in based application is convenient to process the expected high throughput data flow in parallel at next-generation facilities such as the High Energy Photon Source (HEPS). The objectives, functionality and architecture of Daisy are described in this article.

## 1 Introduction

Large scale research facilities are becoming prevalent in the modern scientific landscape. One of these facilities' primary responsibilities is to make sure that users can process and analyze measurement data for publication. To allowing barrier-less access to those highly complex experiments of a broad multi-disciplinary user, nowadays, almost all beamlines require fast feedback capable of manipulating and visualizing data on-line to offer convenience for the decision process of experimental strategy. And recently, the advent of beamlines at fourth-generation synchrotron sources and the advanced detector has made significant progress that push the demanding of computing resource at the edge of current workstation capabilities. On top of this, most synchrotron light source has shifted to prolonged remote operation because of the outbreak of a global pandemic, with the need for remote access to the on-line instrumental system during the experiments. Another issue is the vast data volume produced by specific experiments, making it difficult for users to take data away. In this case, as shown in Fig.1, on-site data analysis services are necessary both during and after experiments.

---

[*] Corresponding author: tianhl@ihep.ac.cn
[#] Corresponding author: qfz@ihep.ac.cn

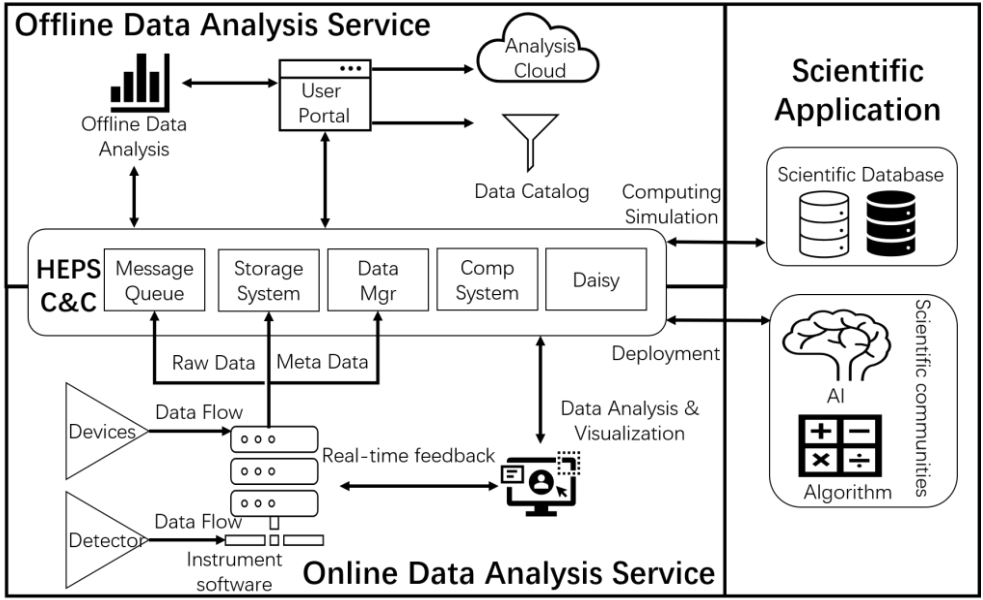

**Fig. 1.** The integrated computing platform of the HEPS
enable on-site data analysis services both for online and offline

The High Energy Photon Source (HEPS) [1] is the first high-performance fourth-generation synchrotron radiation light source with a diffraction-limited storage ring in China. It creates opportunities for researchers to perform multi-dimensional in-situ characterization of complex structures and dynamic processes with a high beam energy of 6GeV and ultra-low emittance of 0.06 nm*rad. Some state-of-the-art experimental techniques, such as phase-contrast tomography and ptychography approaches, will become commonplace. Customized algorithms developed in the scientific communities are needed to take these techniques' full advantages.

However, it raises a critical problem of integrating this algorithmic development into a novel computing environment that can be used in the experimental workflow [2]. The solution requires collaboration with the user research groups, instrument scientists and computational scientists. A unified software platform that provides an integrated working environment with generic functional modules and services is necessary to meet these requirements [3,4,5]. Scientists can work on their ideas, implement the prototype and check the results following some conventions without suffering from the technical details and the migration between different HPC environments. Thus, one of the vital considerations is the capability of flexible and configurable integrating extensions to existing software. Another challenge resides in the interactions between instrumental sub-systems, such as control system, data acquisition system, computing infrastructures, data management system [6], data storage system and so on, which can be quite complicated.

The Daisy (Data Analysis Integrated Software System), based on an object-oriented plug-in architecture, has been designed to address these challenges. The project aims to bridge the gap with an all-encompassing framework between sophisticated computing infrastructure and the users focusing on scientific issues. This article describes the main components of this software and the status of the progress.

## 2 Design

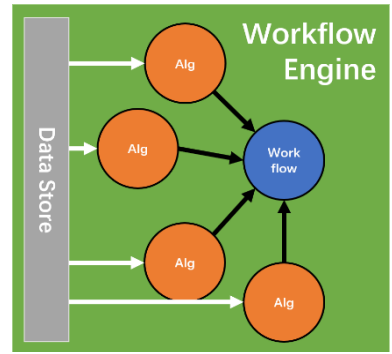

Fig. 2. Daisy framework design

The main goals of Daisy are the following:

1. Provision of an extensible framework that can easily port existing algorithms.
2. Scientific workflow that can involve reusable algorithms in parallel computing system.
3. Integrated capabilities that can adopted from other application.
4. Data visualization and analysis services that can access via a browser and native application

As shown in Fig.2, the Daisy framework consists of a highly modular C++/Python architecture which is composed of four pillars: datastore, algorithm, workflow and workflow engine. The primary design consideration is the separation of data object and algorithms [7]. Datastore is the data object container that passes on data objects between algorithm. And the algorithm, the handler of existing 3rd codes, gets input data objects from Datastore, passes them out to the external libraries, generates output one and sets them back to Datastore. The workflow, a sequence of algorithms defined by scientists, is scheduled by the workflow engine. The architecture that supports customized plug-in functions can easily access visualization tools and the Python scientific computing ecosystem. The graphic user interface employs MVC architecture for handling data objects, analysis workflow and presentation. Web and native application based on Jupyter notebook [8] and PyQt [9] integrate script editor, data visualizing and processing widgets.

Daisy is designed to separate two main layers: (1) a scientific domain layer focusing on the algorithm, workflow and user interface, and (2) an IT infrastructure layer that interacts with the computing environment and other instrumental sub-systems. The scientific domain layer considers three types of scientists: less experienced users who only wishes to execute the pre-defined workflow through the graphical interface or command-line interface; instrument scientists and experts who implement the workflow, specify its parameters, and develop customizable interface; and data scientists who study the data processing algorithms and tools for analysis task. And the IT infrastructure layer, including Datastore and Workflow Engine, are developed and tested by software engineers. This stratification allows the separation of the technological details and the implementation of calculation. This generic and modular workflow structures can support an extensive range of scientific applications.

### 2.1 Algorithm, Data Object, and workflow

A Daisy process relies on a Workflow instance to execute a sequence of Algorithm instances. The interaction with the Daisy framework occurs through the Python-based Application Program Interface (API) and the Graphical User Interface (GUI). The Algorithm module with pre-defined methods can be initialized, configured, executed and finalized by the Workflow module, whose responsibility is performing a particular action, for example, invoking an external library or a calculation task. To ensure performance as well as flexibility, Python/C++ are both supported to implement the Algorithm module. The Python Algorithm class is derived from the C++ one, and the framework provides automated python bindings via Boost.Python for C++ Algorithm, so those two types of class perform in the same behaviours pattern in the Python environment.

The Data Object managed by the Data Store module is an in-memory transient data structure that facilitates information exchange between algorithms. They can be loaded from various file formats, live streams, or created by algorithms. Except for the Persistence Algorithms (e.g. LoadAlg and SaveAlg), ordinary algorithms should not access the persistency store but instead use the Data Objects registered in the Data Store module. In order to combine multiple scientific software packages where the data structure will be quite complex and unpredictable, the framework does not provide the unitary Data Model. Still, it compulsorily requires NumPy numerical types as its basic data structures acceptable by Python and C++ execution modules.

In the practical term, the Algorithm and the Data Object are both managed by the global mapping relation between the name and the instance, and the details will be described in following section. The Workflow module thus connects the sequence of Algorithms to be executed in order, and coordinates the flow of Data Object between Algorithm. The workflow also keeps a set of parameters given by users, and holds a set of properties that required by algorithms. Since a workflow containing a set of properties can act in the same behaviours as an algorithm to perform the execution task, the workflow class is driven by the algorithm class. It is recommended to organize a complex workflow in a tree structure: a group of algorithms and sub-workflow (optional) embodies a workflow.

The framework needs customization to implement different data processing application in various environments. Usually those scientific specific codes will eventually be encapsulated in two places: Algorithm and Workflow. With a set of well-defined functions for certain type of interaction, they both support run-time loading via dynamic libraries, and allow us to implement true plug-and-play application. The examples of those two modules sharing similar interfaces can be achieved, as shown in Fig.3.

```
import numpy as np                                    @Daisy.Singleton
import tomopy                         A               class WorkflowCTReconstruct(Daisy.PyWorkflow):      B
from Daisy import DaisyAlg                                def execute(self):
class AlgTomopyRecon(DaisyAlg):                               self.engine['loadhdf5'].execute(input_path='/entry/tomo', output_dataobj='tomodata')
    def __init__(self, name):                                 self.engine['loadhdf5'].execute(input_path='/entry/dark', output_dataobj='darkdata')
        super().__init__(name)                                self.engine['loadhdf5'].execute(input_path='/entry/flat', output_dataobj='flatdata')
                                                              self.engine['normalize'].execute(projs_dataobj='tomodata', darks_dataobj='darkdata',\
    def initialize(self):                                                      flats_dataobj='flatdata', output_dataobj='normdata')
        self.data = self.get("DataStore").data()              self.engine['angles'].execute(input_dataobj='normdata', output_dataobj='thetas')
        self.LogInfo("initialized, Tomopy Reconstruction")    self.engine['minuslog'].execute(input_dataobj='normdata', output_dataobj='mlogdata')
        return True                                           self.engine['reconstruct'].execute(input_dataobj='mlogdata', theta='thetas',\
                                                                              center=1030, alg_type='fbp',output_dataobj='recodata')
    def execute(self, input_dataobj, theta, center, alg_type, output_dataobj):
        projs = self.data[input_dataobj]                      self.engine['savehdf5'].execute(input_dataobj='recodata',output_path='/entry/reco')
        thetas = self.data[theta]
        dataobj = tomopy.recon(projs, thetas, center=center, algorithm=alg_type)
        self.data[output_dataobj] = dataobj            wf = workflowCTReconstruct('WorkflowCTReconstruct')
        return True                                    wf.initialize(workflow_engine='PyWorkflowEngine', \
                                                                    workflow_environment = init_dict, algorithms_cfg = cfg_dict)
    def finalize(self):                                wf.execute()
        self.LogInfo("finalized")
        return True                                    data =wf.data_keys()
                                                       algs =wf.algorithm_keys()

                                                       wf.finalize()
```

**Fig. 3.** Examples of (A) how to invoke an external software within the Algorithm module, and (B) how to implement and execute a Workflow module

## 2.2 Workflow Engine and Data Store

The data centring architecture is implemented through the Data Store to manage the data propagation, which is an intermediate buffer to minimizing the coupling between Algorithms. In the execution process, the Data Object generated by the Algorithm registers into the Data Store and can be achieved by other Algorithm in the workflow. The Data Store initializes automatically by the Workflow Engine before the execution of the Worklfow and can be accessed by every Algorithm (as shown in Fig.2 A, initialize function). The Algorithm requires Data Object through the Data Store through its name, which is compulsory unique in the workflow. If the new Data Object is regretted in the same name as an older one, the Data Store will replace the expired one with the new one. In this process, there is no need for the user to considerate memory management, which is a trick, especially on the distributed system. As shown in Fig.3, the Data Object's name is defined in the workflow, by which the Algorthm can access the Data Object's instance in the Data Store. This approach is adopted by Gaudi and later Mantid software. Considering the development of computing techniques, in Daisy, however, the Data Store is provided as a backend that can be defined by the Workflow Engine. This mechanism ensures the migration of the workflow from one platform to another and preserves the heritage of scientific communities.

```
init_dict ={'loaddata':{'class_name':'LoadHDF5',\
                        'init_paras':{'inputfile_name':'scan_00575_data_000001.h5'}\
                       },\
            'loadmask':{'class_name':'LoadHDF5',\
                        'init_paras':{'inputfile_name':'scan_00575_master.h5'}\
                       },\
            'azintalg':{'class_name':'AlgFAIIntegrate',\
                        'init_paras':{'wavelength':'0.7293'}\
                       },\
            'savedata':{'class_name':'SaveHDF5',\
                        'init_paras':{'outputfile_name':'test_scan.h5'}\
                       }\
           }
```
A

```
cfg_dict = {'azintalg':{'directDist':169,\
                        'centerX':1049.967,\
                        'centerY':1063.892,\
                        'pixelX':75,\
                        'pixelY':75,\
                        'PlanRotation':64.66877,\
                        'tilt':0.3753,\
                        'azimin':93,\
                        'azimax':115,\
                        'radmin':10.0,\
                        'radmax':20,\
                        'ntth':40}\
           }
```
B

**Fig. 4.** Examples of workflow configurations

As shown in Fig.3 B, The Workflow Engine schedules the Algorithms explicitly through their names defined in the workflow. The Workflow Engine, the intermedia layer between the computing environments and the scientific application, are responsible for the initialization of the Data Store and the Algorithm, sequential or parallel execution of Algorithms, and other essential services such as logging and error tracking. Computing resources and initial parameters need to specify in the beginning phase of workflow execution. As shown in Fig4, two types of JSON scheme are presented here. Fig.4.A shows the requirements of the module, which includes the module name, the version, and other information required by the Computing Infrastructure. And Fig.4.B shows the parameters needed by the algorithms, which are usually stored in the Data Management System and maintained by instrument scientists.

The foundation of the in-process Workflow Engine is an in-house developed framework, SNiPER [10], which supports many experiments led by IHEP, like JUNO [11], LHAASO [12], and CSNS [13]. Its light-weight plugs-in architecture inspired by the GAUDI [7] framework uses a key-value dictionary as the index for locating and invoking the dynamic library across the C++ and Python modules. Furthermore, as shown in Fig.5, the distributed Workflow Engine is based on Apache Spark [14], a unified analytics engine for large-scale data processing. Firstly, the Spark Workflow Engine creates SparkContext, which employs the Resilient Distributed Dataset (RDD) as its Data Store. Secondly, the workflow is split into many small Spark tasks (sub-workflow) distributed to worker nodes. The SNiPER workflow engines on the worker nodes achieve data from RDD and execute sub-workflow simultaneously. Finally, the output data generated by sub-workflow are collected in RDD and sent back to Spark Driver Program (top-workflow). In this case, the Spark workflow engine translates the top-workflow script into multiple sub-workflow executed by the SNiPER workflow engine on the cluster.

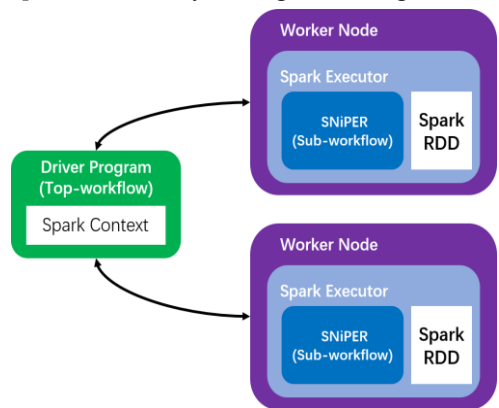

**Fig. 5.** Workflow on Apache Spark

## 3 Prototype: HEPS testbed at 3W1A of BSRF

Beijing Synchrotron Radiation Facility (BSRF) [15] sets up a testbed at beamline 3W1A dedicated to verifying state-of-the-art techniques. As shown in Fig.6.A, The HEPS computing and communication group (HEPS-CC) also establish an integrated computing environment providing the advanced information infrastructure for high-throughput analysis services. The Data Management System [6], based on SciCat [16], aims to manage the whole data lifecycle. The Data Storage System employs Lustre [17] to satisfy the requirements of massive data production in the HEPS. The Computing System focuses on the Kubernetes [18], Docker [19] and JupyterHub [20] to implement a scalable and flexible computing environment for users. Mamba, a science-oriented data acquisition software on top of Bluesky [21], offers a highly customizable solution for experimental plans and high-performance data multiplexing, as well as the capabilities of natural integration with Daisy.

Integrating an open-source 3D tomographic reconstruction package, TomoPy [22], into the framework, Daisy provides a remote accessing client employing Jupyter message protocol and data visualization based on Jupyter Widget (as shown in Fig.6.B). Mamba trigs the pre-defined workflow script, which is automatically scheduled by a Kubernetes cluster. The parallelization supported by Spark cluster and incorporation with DMS is on the way.

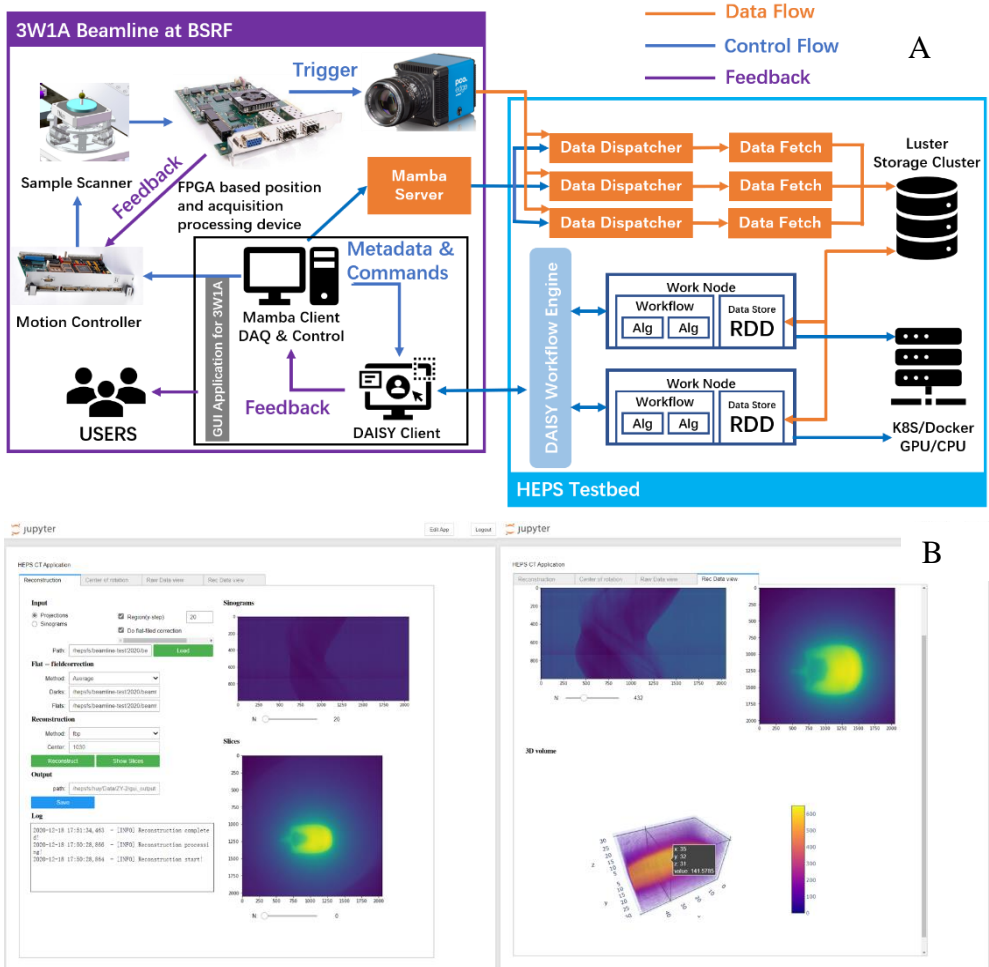

**Fig. 6.** (A)The scheme of HEPS testbed for tomographic experiment at 3W1A beamline of BSRF; (B)The Web-based graphic user interface powered by Jupyter Widget.

## 4 Status and Outlook

The proof-of-concept design has been implemented, which is now being used to perform tomography reconstruction application on the HEPS testbed. Development is proceeding by expanding an extensive range of necessary components, including algorithms that integrate the TomoPy and HDF IO [23] module h5py [24] and data visualization service based on Jupyter Widget. In the area of the distributed computing system, the development of Spark Workflow Engine is continuing that involves PySpark and HDF5 connector [25] for Apache Spark. In addition, the user interface is also extended to support PyQt5 and provide remote access through the Jupyter message protocol. Furthermore, incorporations with scientific communities are also essential to work in this development. A user-friendly integrated development environment and graphic user interface should be provided for end-users to integrate algorithms and set up data analysis applications.